# A Smart Cushion for Real-Time Heart Rate Monitoring

Chacko John Deepu[1], Zhihao Chen[2], Ju Teng Teo[2], Soon Huat Ng[2], Xiefeng Yang[2], and Yong Lian[1]
[2]Institute for Infocomm Research, Agency for Science, Technology and Research, Singapore.
[1]Department of Electrical and Computer Engineering, National University of Singapore, Singapore
Email: eleliany@nus.edu.sg

*Abstract*— **This paper presents a smart cushion for real time heart rate monitoring. The cushion comprises of an integrated micro-bending fiber sensor, which records the BCG (Ballistocardiogram) signal without direct skin-electrode contact, and an optical transceiver that does signal amplification, digitization, and pre-filtering. To remove the artifacts and extract heart rate from BCG signal, a computationally efficient heart rate detection algorithm is developed. The system doesn't require any pre-training and is highly responsive with the outputs updated every 3 sec and initial response within first 10 sec. Tests conducted on human subjects show the detected heart rate closely matches the one from a commercial $SpO_2$ device.**

## I. Introduction

An estimated 70 million people in the United States suffer from sleep problems, and more than 50% of them have a chronic sleep disorder. Each year, sleep disorders, sleep deprivation, and excessive daytime sleepiness add approximately $16 billion annually to the cost of health care in the U.S., and result in $50 billion annually in lost productivity. In the elderly, it is estimated that about 25% suffer from some form of sleep disorder [1]. It is known that sleep deprivation leads to a deterioration in work performance and productivity. Epidemiologic studies have found an association between cardiovascular morbidity and chronic sleep restriction. In the Nurses' Health Study, women sleeping less than 7 hours a night had increased risk of coronary events compared to those averaging 8 hours of sleep a night [1]. Hence, there is a need to develop a non-intrusive device for monitoring sleep quality.

Research has shown that heart rate variability (HRV) is a good indicator for sleep quality [2]. The major step in measuring HRV is to compute instantaneous heart rate from patient's (ECG) electrocardiogram. However, this requires the attachment of several electrodes to skin, which is not convenient for home usage. One possible alternative is to use BCG for heart rate detection. BCG is a technique to record body vibrations caused by the activity of the heart. It could be used for cardiac evaluation and sleep monitoring [3-4]. The advantage of BCG over traditional ECG is its non-contact nature, i.e. there is no skin-electrode contact during the measurement. Thus it is well suited for sleep quality monitoring. In addition, it can also be used for other applications such as driver fatigue detection, heart rate detection during MRI scanning.

There are a few research activities on various types of BCG sensor including a force sensitive electromechanical film (EMFi) sensor [5], weighing scales [6], pressure pad and load cell [7-8]. These sensors are electronic based and susceptible to electromagnetic interference, which may be a problem in some clinical applications, e.g., Medical Resonance Imaging (MRI). Fiber optic sensor is a better candidate in medical applications because of its inherent immune to electromagnetic interference. Fiber Bragg grating sensor based on wavelength modulation [14] is reported as a potential candidate. Also, a micro-bending fiber based BCG sensor was reported in [9], in which a threshold algorithm is used to calculate the heart rate. The threshold algorithm seems to be not robust, e.g. its accuracy is easily affected by motion artifact. In this paper, we propose a smart cushion for heart rate monitoring. The cushion consists of an integrated micro-bending fiber sensor [9] for acquiring BCG signal, and a new heart rate extraction algorithm. The proposed setup and algorithm doesn't require any pre-training and is highly responsive with the outputs updated every 3 sec and initial response within first 10 sec.

The paper is organized as follows. Section II introduces the system architecture. Section III presents the heart rate extraction algorithm. The test results are given in Section IV. The conclusion is drawn in Section V.

## II. System Architecture

The principle of using micro-bending fiber sensor for heart rate measurement is based on BCG, which measures the body vibration caused by the heartbeat. A highly sensitive micro-bending fiber sensor [9] is adopted in our system to acquire BCG signals. The detection of vibration is based on fiber optic micro-bending theory. The sensor contains a section of graded multimode fiber clamped between a pair of micro-benders as shown in Figure 1. As the displacement between two micro-benders changes, the light intensity of the clamped multimode fiber changes with subject's body vibrations caused by respiration/heart beating, i.e. the light intensity in the micro-bending fiber is modulated by the body vibrations. This modulated signal is picked up as BCG signal by a detector inside the optical transceiver.

The transceiver contains a light source, a light detector, amplifiers, filters, an analog-to-digital converter, a microprocessor, and some interface circuits for connecting

This work was supported by the Agency for Science, Technology and Research (A*STAR) Biomedical Engineering Programme Grant 102 148 0011.

the transceiver to a computer via USB cable. A band-pass filter is built into the transceiver to remove the low frequency respiratory motion and high frequency noise. Although BCG signal spectrum, ranges from 0.5-20Hz, we used a higher cutoff frequency (of 1.6Hz) to filter out the respiratory movements from the signal. Further, the BCG signal is sampled at 50Hz.

The block diagram of the smart cushion is shown in Figure 2. The micro-bending sensor is embedded inside a cushion as highlighted in blue. The sensor mat is made of micro-bender, multimode fiber that covers whole cushion, and cover materials.

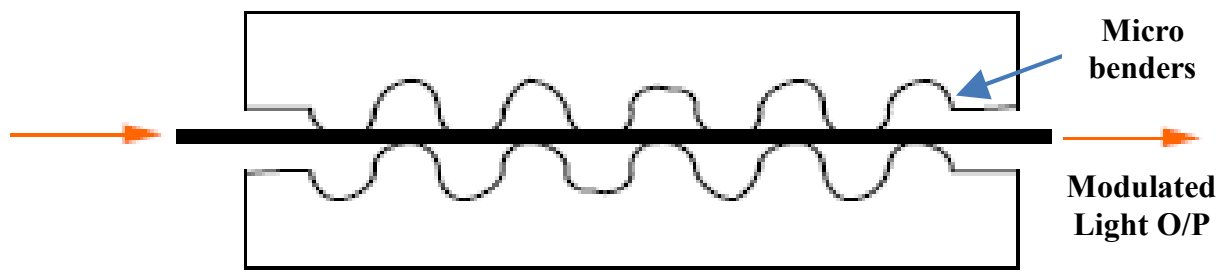

Fig. 1 Micro-bending fiber sensor.

During the heart rate measurement, the cushion can be placed either on a chair, as shown in Figure 3, or be used as a pillow. The back (used for measurements) or head of a person is in contact with the cushion. The cardiac beating induces body movements on the cushion. This in turn produces a loss change in light intensity that indicates cardiac beating. The variation in the light intensity is picked up by an optical transceiver and processed by the proposed algorithm.

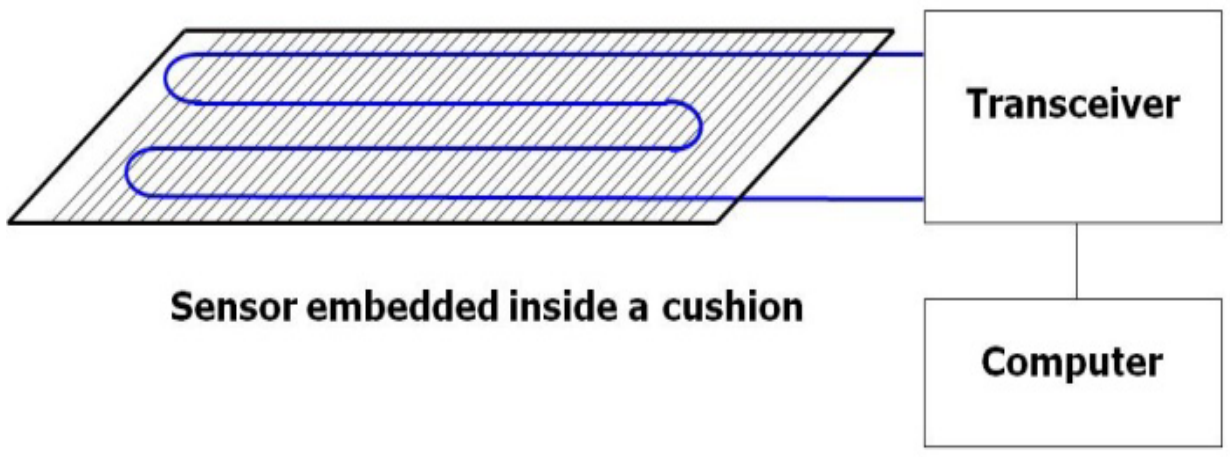

Fig.2 Block diagram of smart cushion.

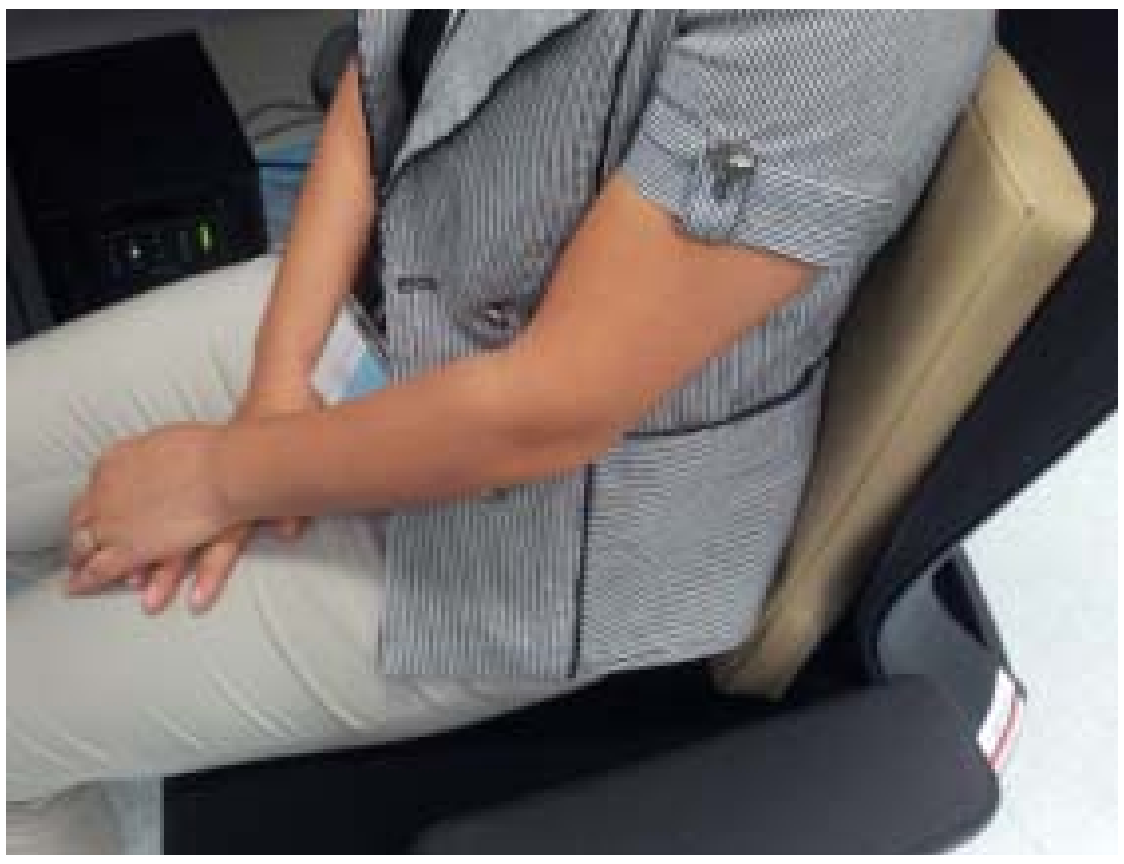

Fig. 3 The use of smart cushion.

A typical BCG waveform acquired from a healthy subject using the proposed smart cushion is shown in Figure 4. It is clear that the acquired BCG signal closely resembles the BCG waveform given in other references [3]. The main components of BCG, e.g. H, I, J, K, L, & N waves, can be clearly identified.

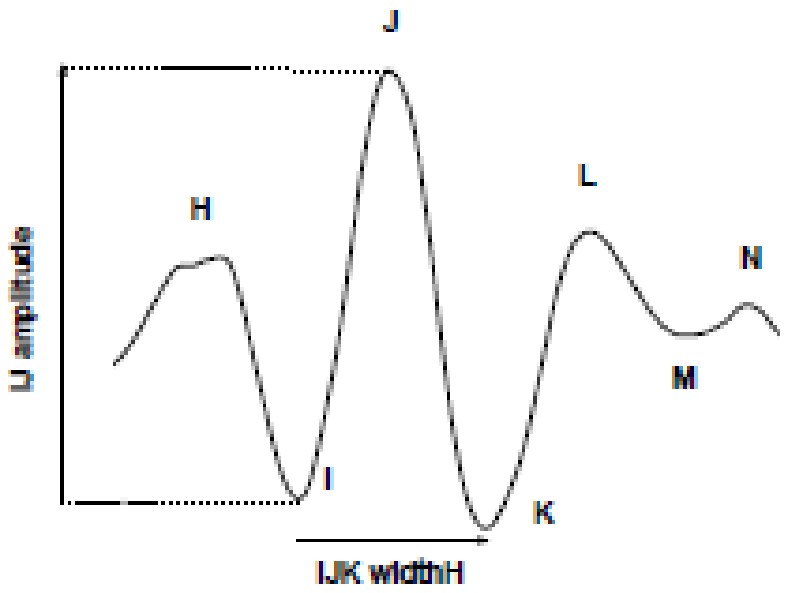

Fig. 4 BCG signal acquired from a healthy subject using the proposed system.

## III. THE PROPOSED HEART RATE EXTRACTION ALGORITHM.

There are several attempts [11-13] to detect heartbeats from BCG signal. A wavelet transform technique was introduced in [11] for the detection of heart rate. Machine learning for searching BCG peaks was reported in [12]. An exhaustive template correlation algorithm for searching the BCG peaks was presented in [13]. These algorithms gave reasonable detection accuracy. However the computational complexity may be a concern if these algorithms are used in portable battery operated device. Our aim is to develop a low complexity heartbeat detection algorithm suitable for ASIC implementation.

The BCG signal obtained from the smart cushion shows distinctive peaks, as shown in Figure 5, which correspond to the ejection of blood into the vessels with each heartbeat. Along with these peaks, there are low frequency, time varying noise and high frequency impulse noise. The spectrum of the BCG signal, as shown in Figure 6, acquired from several subjects was analyzed to understand the frequency components that correspond to the peaks in the BCG signal. It was found that main spectral components in the BCG signal are located in 2-10 Hz frequency range (marked as the red box), which is a subset of the range is specified in [10]. Based on these observations, a heartbeat detection algorithm was developed to extract the heart rate from the acquired BCG signals.

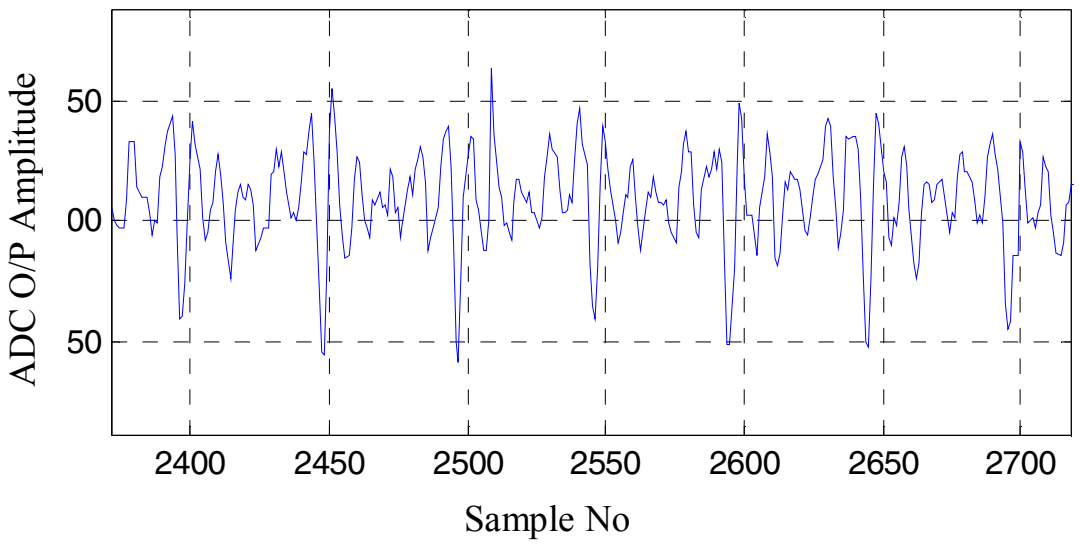

Fig. 5 BCG Signal from smart cushion.

The proposed algorithm includes 6 steps, as shown in Figure 7. The BCG samples are filtered using a 2-10 Hz band-pass FIR filter with a 40dB attenuation to remove the low and high frequency impulse noises. The amplitude swing of the resultant signal is enhanced by conducting a cubing ($x^3$) operation on the filtered signal while keeping the signal sign intact. In the next step, a moving average operation is performed over 0.06 second (3 samples), in order to filter out any momentary upswing/downswing. The width of the BCG IJK complex is greater than this duration and hence is not affected. The filtered output is further smoothened by computing the absolute value and averaging over 0.3 second (15 samples) before doing a cone detection and comparing to a detection threshold. The outputs from various stages are illustrated in Figure 10.

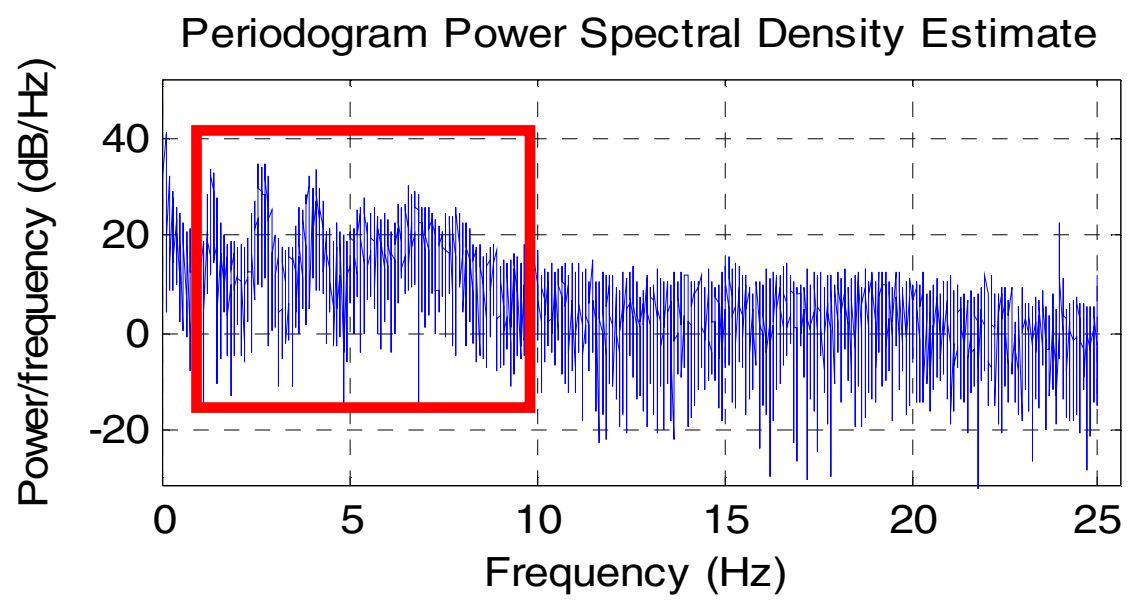

Fig. 6 Frequency spectrum of BCG signal.

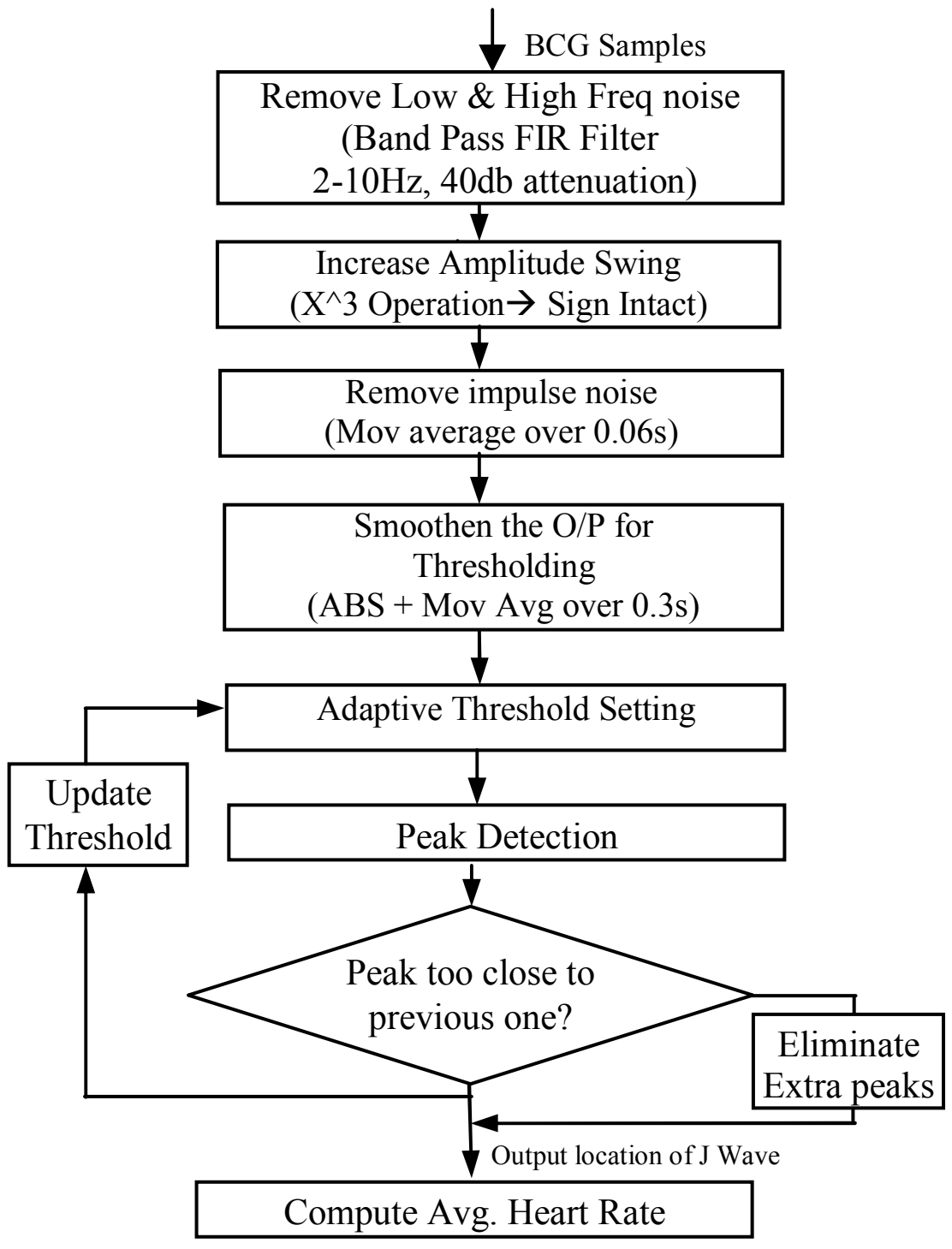

Fig. 7 BCG heart rate detection flow chart.

The performance of proposed algorithm is dependent on the threshold setting. This is because the filtered signal is constantly compared against the threshold to check for the presence of the peak. To improve the accuracy, we proposed an adaptive threshold setting method. The threshold value is initialized as percentage of the maximum value of the initial 300 samples. Every time the smoothened signal exceeds the threshold, the peak detection algorithm searches for the presence of a peak, as described later. The threshold is computed as 25% of the average of last 8 detected peaks. In order to prevent sudden amplitude variations drastically affecting the threshold computation, the maximum peak amplitude considered for each detection is limited to a maximum of 2 times that of the previously detected peaks.

To prevent a high threshold causing a lockup, in which all subsequent peaks are not detected, an automatic threshold reduction in a step of 10% of previous peak is implemented for every J-J interval, where no detections are made. An average of past 8 successful detections are used to calculate the J-J interval used above. The automated threshold adaptation for BCG is illustrated in Figure 8.

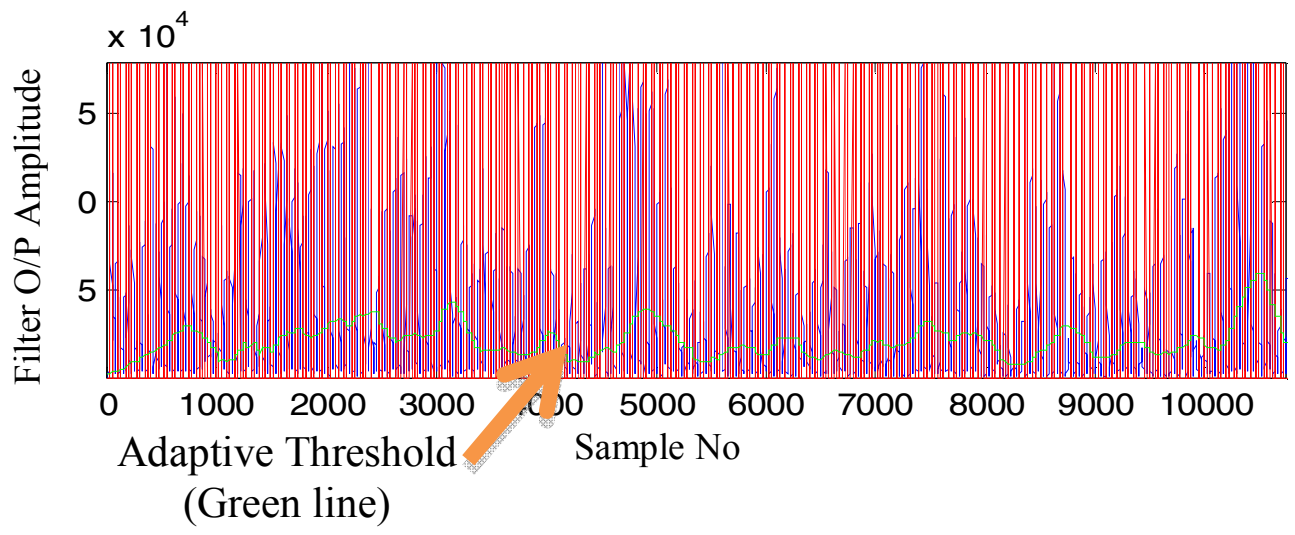

Fig. 8 Adaptive threshold setting.

The peak detection algorithm kicks in when the filtered signal exceeds the threshold. The detection starts with finding a constantly rising edge and then a continually falling edge within a specific period of time. The rising edge is identified by using the following conditions:

$$x(i) - x(i-1) > 0\,, \quad x(i-1) - x(i-2) > 0, \qquad (1)$$

$$x(i-2) - x(i-3) > 0, \quad x(i-3) - x(i-4) > 0, \qquad (2)$$

where $x(i)$ is the $i^{th}$ sample of smoothened moving average output. After a rising edge is located, the algorithm looks for a falling edge within a search window of 0.25 second (12 samples). A falling edge is determined using the following conditions, and if present, a peak is declared as shown in Figure 9.

$$x(i+1) - x(i) < 0\,, \qquad (3)$$

$$x(i+2) - x(i+1) < 0, \qquad (4)$$

$$x(i+3) - x(i+2) < 0, \qquad (5)$$

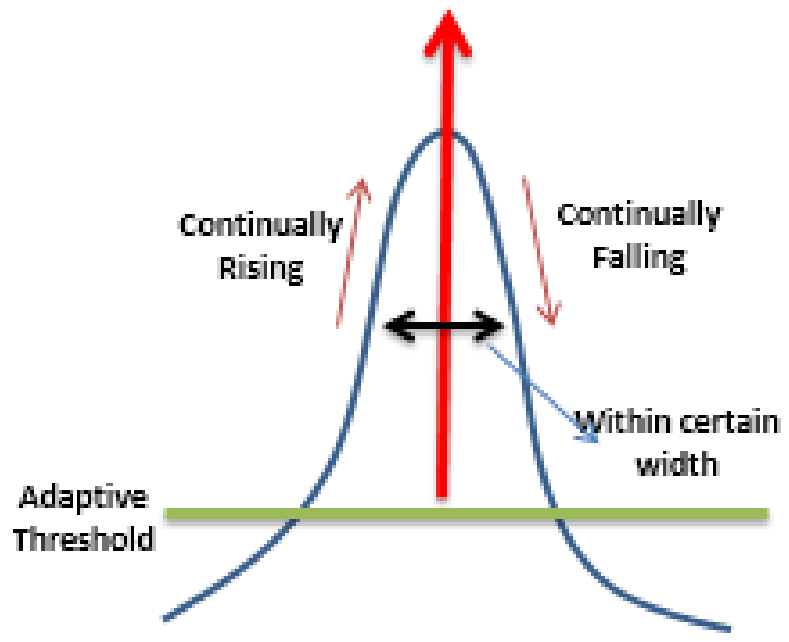

Fig. 9 Peak detection.

If a peak detected is very near, e.g. within 0.3 second, to a previously detected peak, then it is most likely, that one of them is a false detection. To avoid this scenario, when detected peaks are too close, one of them is discarded, i.e. the one with lower peak amplitude.

`

## IV. TEST RESULTS.

The smart cushion and proposed algorithm is tested on 5 volunteers, both male and female, with heights ranging from 154-172 cms, weights ranging from 53-68 Kgs, and age from 25-49 years, while at rest. A commercial pulse oximetry device was also used simultaneously to capture the heart rate from the subjects for 5 minutes/subject. To evaluate the performance of heart rate detection, false positive (*FP*) and false negative (*FN*) are observed from the algorithm outputs. False positive means a false beat is detected when there is none, and false negative means that the algorithm fails to detect a true beat. Furthermore, by using *FP* and *FN*, the sensitivity (*Se*), positive prediction (+*P*) and detection error (*DER*) are calculated using the following equations:

$$Se\ (\%) = \frac{TP}{TP + FN}, \quad (6)$$

$$+P\ (\%) = \frac{TP}{TP + FP}, \quad (7)$$

$$DER = \frac{FP + FN}{Total\ Peaks}, \quad (8)$$

where *TP* stands for true positive, i.e. the number of peaks correctly detected. Tables I and II show the performance of detection algorithm.

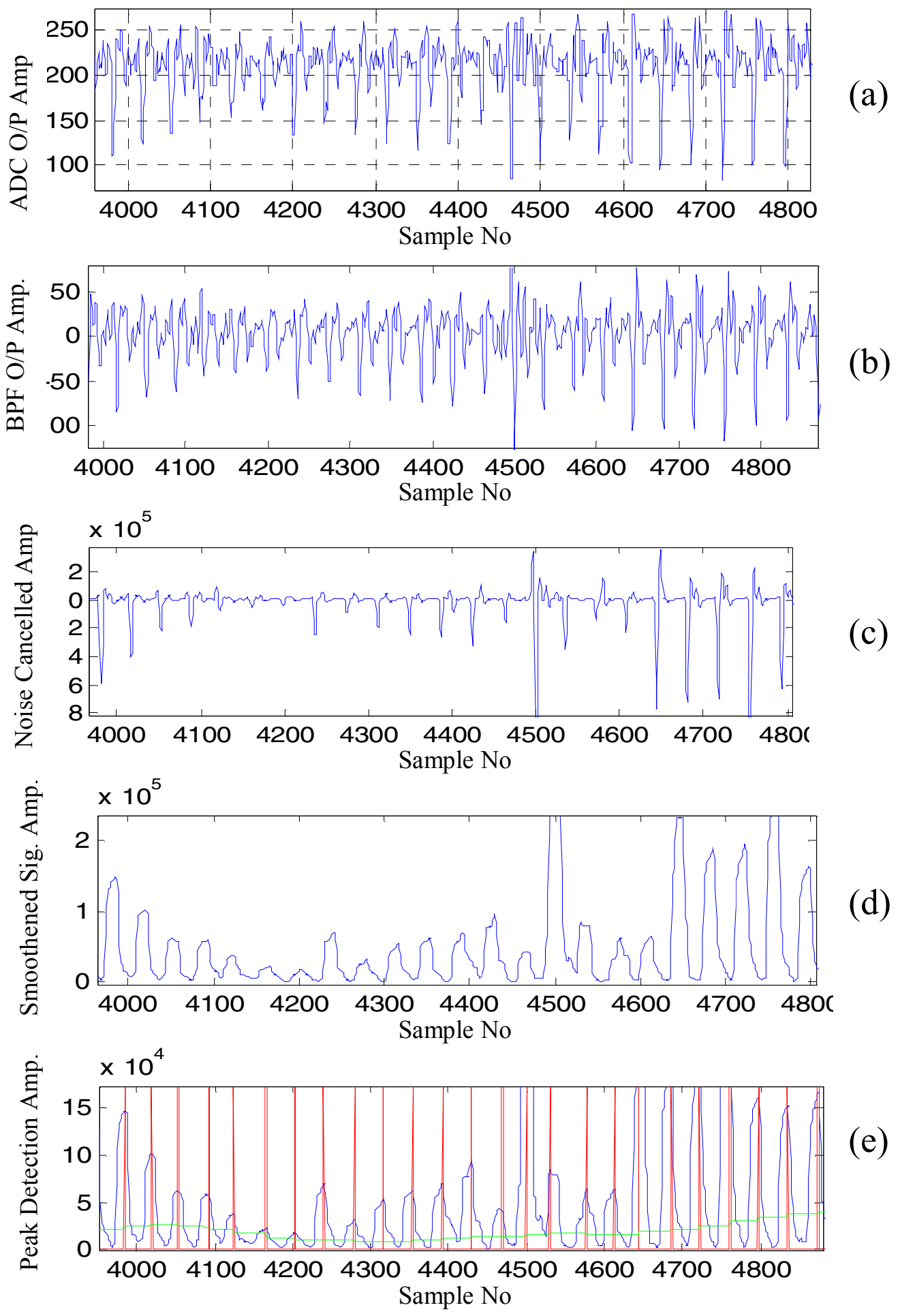

Fig. 10 Outputs in BCG signal processing: (a) original input; (b) band-pass filtered signal; (c) after $x^3$ and moving average; (d) absolute value and moving average; (e) after peak detection.

TABLE I
HEART BEAT DETECTION PERFORMANCE OF THE PROPOSED SYSTEM

| User | Total | FP | FN | Se (%) | +P (%) | DER |
|---|---|---|---|---|---|---|
| I | 124 | 3 | 7 | 94.35% | 97.5% | 0.080 |
| II | 306 | 0 | 4 | 98.69% | 100.00% | 0.013 |
| III | 302 | 1 | 6 | 98.01% | 99.66% | 0.023 |
| IV | 166 | 6 | 5 | 96.98% | 96.40% | 0.066 |
| V | 215 | 4 | 7 | 96.74% | 98.11% | 0.051 |

TABLE II
RESULTS COMPARISON

| User | Average Heart Rate by Pulse Oximetry | Heart Rate computed using proposed system |
|---|---|---|
| I | 66 | 61.3 |
| II | 77.5 | 78.4 |
| III | 74.5 | 73.5 |
| IV | 59 | 63.4 |
| V | 67.5 | 67.2 |

## V. CONCLUSION

We have presented a micro-bending optic sensor based smart cushion for non-intrusive heart-rate monitoring as well as a heart rate detection algorithm for BCG signal. Tests conducted on 5 volunteers show that the proposed system achieves an accuracy of above 95% on an average.